\documentclass[aps,prb,twocolumn,groupedaddress,amsmath,amssymb]{revtex4-1}
\usepackage{graphicx}  
\usepackage{dcolumn}   
\usepackage{bm}        
\usepackage{verbatim}   

\begin{document}

\title{Coherent Transport of Levitons Through the Kondo Resonance}
\author{Takafumi J. Suzuki}
\altaffiliation[Present address: ]{
  Department of Applied Physics, Graduate School of Engineering, The University of Tokyo, Japan
}
\affiliation{
  Department of Physics, The University of Tokyo, 7-3-1 Hongo, Bunkyo-ku, Tokyo 113-0033, Japan
}
   
\date{\today}

\begin{abstract}

We study coherent transport of levitons through a single-level quantum dot system driven by Lorentzian-shaped voltage pulses.
We demonstrate the repeated emergence of the Kondo resonance in the dynamical regimes where the Fermi sea is driven by optimal pulses free of particle-hole excitations.
The formation of the Kondo resonance significantly enhances the dc transport of levitons.

\end{abstract}

\maketitle


The Kondo effect has been one of the central subjects of condensed matter physics over the past 50 years \cite{kondo1964resistance,hewson1997kondo}.
It is an archetypal example of coherent many-body phenomena in interacting electronic systems and essential low-energy behaviors of Kondo systems are described by the local Fermi liquid theory.
The development of nanotechnology has extended the Kondo physics to nonequilibrium regimes \cite{Goldhaber-Gordon1998,*Cronenwett540,*schmid1998quantum,vanderWiel2105,PhysRevLett.106.176601,Ferrier2016}.
Transport through a quantum dot (QD) enables to study the nonequilibrium Kondo effect with experimentally tuned parameters.
The interplay between the coherent many-body resonance and the nonequilibrium field has posed nontrivial problems.
In particular, determining the fundamental excitation of the Kondo systems driven out of equilibrium remains a challenging topic \cite{doi:10.1143/JPSJ.74.110,PhysRevB.92.075120,Ferrier2016}.
Recently, new insights on the nonequilibrium Kondo physics have been gained in tandem with rapid advances in engineering time-dependent fields.
Electrons dressed with photons acquire novel features which equilibrium electrons do not possess.
Periodic driving fields have been frequently utilized to probe low energy excitations of the QD systems \cite{PhysRevLett.74.4907,PhysRevLett.76.487,PhysRevLett.81.4688,PhysRevLett.81.5394,PhysRevLett.83.808,PhysRevLett.83.384,PhysRevB.61.2146,PhysRevB.64.075319,PhysRevB.81.144301,PhysRevA.94.063647,PhysRevLett.116.026801} and to invent new types of Kondo systems \cite{PhysRevLett.111.215304,PhysRevLett.115.165303}.
In sinusoidally driven QD systems, satellites of the Kondo peak develop at low temperatures due to the absorption and the emission of photons \cite{PhysRevLett.81.5394,PhysRevLett.81.4688}.
On the other hand, the spin-flip cotunneling processes \cite{PhysRevLett.83.384} as well as the ionization of the local site \cite{PhysRevB.61.2146} induce decoherence which hinders the Kondo resonance.

The richness and the complexity of driven electronic systems come from the collective response of electrons under the entire Fermi sea.
In spite of the difficulty originating from the many-body effects, Levitov and coauthors proposed an elegant way to engineer minimal excitation states out of the Fermi sea \cite{levitov1996electron,PhysRevB.56.6839,PhysRevLett.97.116403}.
They found that, among possible pulse profiles, the repeated Lorentzian pulses excite the Fermi sea without creating particle-hole pairs.
The single-particle nature of the excitation was clarified at the same time in terms of the full counting statistics.
The elementary excitations created above the undisturbed Fermi sea are termed levitons, and have been experimentally exploited as ideal fermionic excitations in electron quantum optics \cite{dubois2013minimal,jullien2014quantum}.
There have already been a number of works on levitons injected in quantum Hall systems \cite{PhysRevLett.97.116403,PhysRevB.72.205321,PhysRevB.88.085302,PhysRevB.88.205303,PhysRevLett.113.166403,PhysRevB.94.035404,PhysRevLett.118.076801}.
It is also theoretically proposed that a Fermi sea driven by designed Lorentzian pulses hosts an exotic excitation with a fractional effective charge \cite{PhysRevLett.117.046801}.

In this paper, we demonstrate the coherent transport of levitons through the Kondo resonance in a QD system driven by Lorentzian-shaped periodic pulses.
The Lorentzian driving protocol is distinct from the others because the optimal pulses can excite a fermionic quasiparticle while preserving the structure of the Fermi sea.
This enables the formation of the Kondo resonance under the strong driving field.
We use a many-body approach combined with Floquet's formalism \cite{PhysRevB.78.235124,PhysRevB.91.165302} to provide a conceptually transparent and numerically efficient way to describe the dynamics of the interacting levitons.

%


We consider a single-level QD coupled to left and right leads with a periodically oscillating bias voltage.
The Hamiltonian reads
\begin{align}
\label{eq:Model Hamiltonian}
H = 
& \sum_{\sigma} \epsilon_d
\hat{d}^{\dagger}_\sigma \hat{d}_\sigma 
+ \sum_{\alpha, {\bm k},\sigma} 
\left( \epsilon_{\alpha{\bm k}} + eV_{\alpha}(t)\right)
\hat{c}^{\dagger}_{\alpha{\bm k}\sigma} \hat{c}_{\alpha{\bm k}\sigma} \nonumber \\
& + \sum_{\alpha, {\bm k},\sigma} \left(
t_{\alpha} \hat{d}^{\dagger}_\sigma \hat{c}_{\alpha{\bm k}\sigma} 
+ {\rm h.c.} \right)
+ U \hat{d}^{\dagger}_{\uparrow} \hat{d}_{\uparrow} \hat{d}^{\dagger}_{\downarrow} \hat{d}_{\downarrow} 
,
\end{align}
where $\hat{d}^{\dagger}_{\sigma}$ creates an electron in the QD with spin $\sigma$ and $\hat{c}^{\dagger}_{\alpha{\bm k}\sigma}$ creates a conduction electron in the lead $\alpha$ $(=L,R)$ with spin $\sigma$ and momentum $\bm{k}$.
The coupling $t_{\alpha}$ between the QD and the lead $\alpha$ causes the level broadening
$\Gamma_{\alpha}\equiv 2\pi|t_{\alpha}|^2\rho_{\alpha}$,
where $\rho_{\alpha}$ is the density of states (DOS) of the conduction electrons at the Fermi energy $\epsilon_{F}$.
We consider that the left lead is irradiated with Lorentzian pulses
\begin{align}
  \label{eq: Lorentzian pulses}
  V_{L}(t)=\sum^{\infty}_{m=-\infty}
  \frac{V_{\rm AC}}{\pi} \frac{T_{p}\tau_{w}}{(t-mT_{p})^2+\tau^2_{w}},
\end{align}
with the period $T_{p}$, the width $\tau_{w}$, and the AC amplitude $V_{\rm AC}$.
The right lead is in equilibrium, i.e. $V_{R}(t)=0$.

The periodically driven QD system can be well described in the Floquet-Green's function formalism \cite{PhysRevB.78.235124,PhysRevB.91.165302}.
In the following, we drop the spin index $\sigma$ for simplicity.
The propagators of the photon-dressed interacting electrons are given by the retarded and the lesser Green's functions
$G^{r}(t,t')=-i\theta(t-t')\langle d(t)d^{\dagger}(t')\rangle$
and
$G^{<}(t,t') \equiv i\langle d^{\dagger}(t')d(t)\rangle$,
respectively.
Their Floquet representations are introduced as
$\bm{G}^{r(<)}_{mn}(\omega) \equiv
\int^{\infty}_{-\infty} dt 
\int^{T_{p}/2}_{-T_{p}/2} \frac{dT}{T_{p}}
e^{i\left(\omega+m\hbar\Omega\right)t-i\left(\omega+n\hbar\Omega\right)t'}
G^{r(<)}(t,t')$
with the driving frequency $\Omega=2\pi/T_{\rm P}$.
Hereafter, we use bold letters to denote Floquet-represented functions.

The equilibrium distribution in the right lead is written in the Floquet representation as
${\bm f}_{R}={\bm f}^{\rm eq}$,
where
${\bm f}^{\rm eq}_{mn}(\omega) = \delta_{mn}/(e^{\beta(\omega+m\hbar\Omega-\epsilon_{F})}+1)$
with the inverse temperature $\beta$ and the Kronecker delta $\delta_{mn}$.
In contrast, 
the time-dependent phase 
$\varphi(t) = \frac{e}{\hbar} \int^{t}_{-\infty} V_{L}(t') dt'$
acquired by electrons tunneling from the left lead to the QD significantly modifies the distribution function as
${\bm f}_{L}(\omega) = {\bm U}{\bm f}^{\rm eq}(\omega-eV_{\rm AC}){\bm U}^{\dagger}$.
Here, the dc offset of the periodic Lorentzian pulses is included as a shift of the chemical potential,
and absorption and emission of photons are described via the unitary matrix
$\bm{U}_{mn}=u_{m-n}$ 
with
$u_{l}\equiv \int^{T_P/2}_{-T_p/2} \frac{dt}{T_{p}} e^{i(l\hbar\Omega +eV_{\rm AC}) t} e^{-i\varphi(t)}$.

For the repeated Lorentzian pulses (\ref{eq: Lorentzian pulses}), the matrix elements are computed as
\begin{equation}
  \label{eq: transition probability}
 u_{l}
 =\sum^{\infty}_{k={\rm max}\left\{0,-l\right\}}
 \frac{\Gamma(k +l + q )\Gamma(k - q ) e^{-2\pi\tau (2k+l)} }
      {\Gamma(q) \Gamma(k+l+1)\Gamma(-q) \Gamma(k+1)},
\end{equation}
with 
$q \equiv eV_{\rm AC}/\hbar\Omega$,
$\tau \equiv \tau_{w}/T_p$,
and the Gamma function
$\Gamma(x)$.
The probability to excite $l$-th Fourier harmonics is exponentially suppressed in terms of $l$ due to the factor $e^{-2\pi\tau l}$.
In the delta-pulse limit $\tau \rightarrow 0$, the system is in equilibrium, i.e., $\bm{f}_{L}={\bm f}^{\rm eq}$, at $q \in \mathbb{Z}$
because the phase shift $2\pi q$ introduced by each pulse is exactly absorbed in the gauge degrees of freedom of electrons.
One of the important properties of the quantized Lorentzian pulses is that they generate purely electronic excitations even for finite $\tau$ with minimal disturbance of the Fermi sea \cite{levitov1996electron,PhysRevB.56.6839,PhysRevLett.97.116403}.
It is in contrast to non-quantized pulses which inevitably excite a number of particle-hole pairs as is the case with the Anderson orthogonality catastrophe problem \cite{levitov1996electron,PhysRevB.88.085301}.
These peculiar properties of the quantized Lorentzian pulses result in the quasiparticle nature of levitons created above the undisturbed Fermi sea \cite{PhysRevB.88.085301,PhysRevB.78.245308}.

Since diagrammatic structures of photon-dressed propagators are the same as those in equilibrium, transport properties of interacting electrons are inherited by levitons.
In particular, the Kondo resonance may provide a transport channel of levitons at low temperatures.
However, it is important to note that photon-assisted processes cause two major decoherence mechanisms of the Kondo resonance in driven systems:
(i) reduction of the conduction electrons under the Fermi sea and
(ii) heating effect.
These are crucially dependent on driving pulses via the transition rate $u_{l}$.
Lorentzian pulses are distinct from other driving protocols in that spectral weight is concentrated near the Fermi energy and hole excitations are prohibited at $q\in\mathbb{Z}$.
These favorable natures of the Lorentzian pulses lead to the formation of the Kondo resonance even in nonequilibrium regimes as well as its characteristic dependence on the AC amplitude.
This is the central idea of this paper.

\begin{figure}[b]
\centering
\includegraphics[width=\hsize]{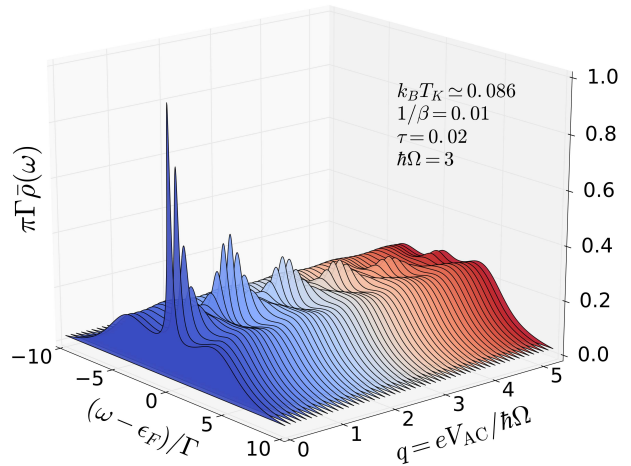}
\caption{
  (Color online)
  Time-averaged DOS
  with
  $\Gamma_{L,R}=1$, $\epsilon_d=-4$, $U=8$, $\beta=100$, $\tau=0.02$, and $\hbar\Omega=3$ for various values of $q$.
  }
\label{fig: DOS}
\end{figure}

\begin{figure*}[t]
\centering
\includegraphics[width=\hsize]{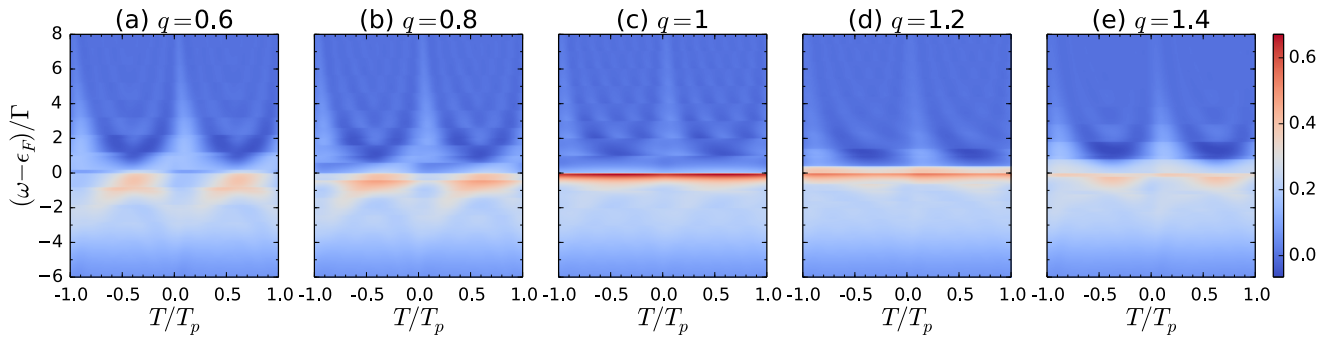}
\caption{
  (Color online)
  Density plots of the Wigner function for various values of $q$.
  The QD system with $\Gamma_{L,R}=1$, $\epsilon_{d}=-3$, $U=6$, and $\beta=100$ is under the periodic Lorentzian pulses with the width $\tau=0.01$ and the driving frequency $\hbar\Omega=2$.
  }
\label{fig: Wigner}
\end{figure*}

One of the hallmarks of the Kondo effect is the appearance of the resonant peak in the time-averaged DOS
\begin{align}
  \label{eq: time-averaged DOS}
  \bar{\rho}(\omega) 
  \equiv 
  -\frac{1}{\pi}  {\rm Im}
  \int _{-\frac{T_p}{2}}^{\frac{T_p}{2}} \frac{dT}{T_p} 
  \int_{-\infty}^{\infty} dt_r 
  e^{i\omega t_r} G^{r}\left(t,t'\right),
\end{align}
with $t_r \equiv t-t'$ and $T\equiv (t+t')/2$.
In this paper, we evaluate the self-energy up to the second order \cite{PhysRevB.46.7046} in $U$ to illustrate qualitative behaviors of the Kondo resonance under periodic Lorentzian pulses.
The second order perturbation theory captures low-temperature Fermi-liquid behaviors, which are of fundamental importance in the leviton transport
\footnote{
A drawback of the $U^2$ approximation is a breakdown of the charge conservation away from particle-hole symmetric cases \cite{PhysRevB.46.7046}.
Systematic treatments of vertex corrections are desirable in future to complementarily confirm the results.
}.
Moreover, it allows us to deal with a large number of ($\sim1/\tau$) Floquet sidebands required to describe photon-assisted processes accompanying a sharp Lorentzian pulse of width $\tau$.
The retarded Green's function can be efficiently calculated in the Floquet representation \cite{PhysRevB.78.235124,PhysRevB.91.165302}
because it has a simple matrix form
${\bm G}^{r}= \left[ {\bm 1} - {\bm g}^{r}{\bm \Sigma}^{r}_{U} \right]^{-1}{\bm g}^r$.
Here, the exact propagators are constructed from the unperturbed one
${\bm g}^{r}_{mn}(\omega)=\delta_{mn}/(\omega+m\hbar\Omega-E_{d}+i\Gamma)$
with
the energy level $E_{d}=\epsilon_{d}+U n_{d}$
and the linewidth $\Gamma=(\Gamma_{L}+\Gamma_{R})/2$.
The charge $n_d$ is determined within the Hartree approximation as 
$n_{d}=-\frac{1}{\pi}{\rm Im}\sum_{m}\int d\omega{\bm g}^{<}_{mm}(\omega)$,
where the unperturbed lesser Green's function is given by 
${\bm g}^{<}={\bm g}^{r} {\bm \Sigma}^{<}_{0}{\bm g}^{a}$
with
${\bm \Sigma}^{<}_{0}=i(\Gamma_{L}{\bm f}_{L}+\Gamma_{R}{\bm f}_{R})$.
The $U^2$ term of the self-energy is given as
$\Sigma_{U}(z,z')=U^2 g(z,z')g(z',z)g(z,z')$
on the Keldysh contour \cite{haug2008quantum}.
The retarded and lesser components of the self-energy are obtained by projecting the Keldysh arguments $z$ and $z'$
onto the real-time axis.

Figure~\ref{fig: DOS} shows the time-averaged DOS for various values of $q$.
We consider a particle-hole symmetric case with $\Gamma_{L,R}=1$, $\epsilon_{d}=-4$, and $U=8$. The temperature  $1/\beta=0.01$ is lower than the Kondo temperature $k_{B}T_{K} \simeq 0.086$ estimated with an expression \cite{PhysRevLett.40.416} 
$k_{B} T_{K}=\sqrt{U\Gamma/2}\exp\left[ \pi \epsilon_{d} (\epsilon_{d}+U ) /2U\Gamma \right]$.
The Lorentzian-shaped bias voltage with $\tau=0.02$ and $\hbar\Omega=3$ excites conduction electrons in the left lead, dressing them with photons.
We use Floquet matrices of size $201$ to describe electrons absorbing or emitting $100$ photons at most.
While the Kondo peak observed at $q=0$ is immediately reduced by the irradiation,
the resonant peak reappears at $q=1$.
The reduction and the formation of the Kondo peak are repeated around the integer values of $q$.
The emergence of the Kondo peak in the dynamical regimes originates from the aforementioned recovery of the Fermi sea around $q \in \mathbb{Z}$:
Electrons in the minimally disturbed Fermi sea form the many-body resonance state with the dot electron.
Away from the optimal points, the driving field produces particle-hole pairs, which inevitably inhibits the Kondo resonance.
The peaks are smeared for $q \gtrsim 1/4\pi\tau$ because the weight of conduction electrons at the Fermi energy decays as $|u_{-n}|^2=e^{-4\pi n\tau}$ at integer values of $q=n$.
The remaining portion of the conduction electrons are distributed above the Fermi energy due to the photon-absorption processes.

The repeated emergence of the Kondo resonance is unfeasible in a sinusoidally driven QD, where the irradiation completely suppresses the fragile many-body resonance in nonequilibrium regimes \cite{PhysRevLett.83.384,PhysRevB.64.075319,PhysRevB.61.2146}.
For the sine-wave case,
the weight of the conduction electrons at $\epsilon_{F}$ decays as $|J_{0}(q)|^2$ with the Bessel function $J_{n}(q)$.
Moreover, disturbance of the Fermi sea by electron-hole excitations causes severe heating effect, which manifests itself as significant enhancement of the current noise \cite{PhysRevB.91.165302}.
These decoherence mechanisms impose severe restrictions on the realization of the Kondo resonance under the sinusoidal driving field.


The dynamical formation of the Kondo resonance can be further analyzed with the electronic Wigner function \cite{PhysRevB.88.205303}
\begin{align}
  \label{eq: Wigner function}
  W(\omega,T) 
  \equiv \frac{\Gamma_{L}\Gamma_{R}}{\Gamma_{L}+\Gamma_{R}}
      {\rm Im} \int_{-\infty}^{\infty} dt_r 
      e^{i\omega t_r}G^{<}\left(t,t'\right).
\end{align}
Here, the lesser Green's function is calculated in the Floquet representation as
${\bm G}^{<}
={\bm G}^{r}\left[{\bm \Sigma}^{<}_{0}+{\bm \Sigma}^{<}_{U}\right] {\bm G}^{a}$.
While it is proportional to a product of the spectral function and the nonequilibrium distribution in stationary cases, photon-assissted processes in driven systems make it difficult to completely separate their contributions.
Nevertheless, the Wigner function is useful to characterize time-resolved excitations
because it visualizes the real-time dynamics of the quasiparticles which are well represented in the energy domain.

The five panels in FIG.~\ref{fig: Wigner} show the electronic Wigner function of the photon-dressed electrons for various values of $q$.
The characteristic structures extending above the Fermi energy are fingerprints of levitons \cite{PhysRevB.88.205303}.
The sharp Lorentzian pulses with $\tau=0.01$ and $\hbar\Omega=2$ create time-resolved excitations centered at $t/T_{p}\in\mathbb{Z}$.
The singularities observed at $\omega-\epsilon_F=n\hbar\Omega/2$ $(n \in \mathbb{Z})$ result from the photon-assisted Fermi edges formed by the driven electrons.
The important difference between the quantum dot system and the previously studied 1D conductors \cite{PhysRevB.88.205303,jullien2014quantum} is the frequency dependence of the spectral function.
At $q=0.6$ [FIG.~\ref{fig: Wigner}(a)], the bare QD level with $\epsilon_{d}=-3$ and $\Gamma_{L,R}=1$ is strongly perturbed by the Lorentzian pulses,
resulting in the decoherence of the Kondo resonance with characteristic energy scale $k_{B}T_{K}/\Delta \simeq 0.16$.
The spectral weight is gradually concentrated at the Fermi energy at a larger pulse amplitude [FIG.~\ref{fig: Wigner}(b)], and eventually becomes a stationary sharp peak at $q=1$ [FIG.~\ref{fig: Wigner}(c)].
In addition to this counterintuitive emergence of the Kondo resonance under the strong driving field,
the coherent ripple patterns
reported in Ref.~\onlinecite{PhysRevB.88.205303}
concurrently become clear at $q=1$.
The quantum interference effects which appear as
negative parts of the Wigner function
may be probed by interferometry and noise measurements \cite{PhysRevB.88.205303,jullien2014quantum}.
When the amplitude becomes larger [FIGs.~\ref{fig: Wigner}(d) and (e)], the Kondo peak is smeared again because of both the dc offset of the pulses and the particle-hole excitations.
The quantum ripples also become obscure away from the optimal situation.



\begin{figure}[t]
\centering
\includegraphics[width=0.9\hsize]{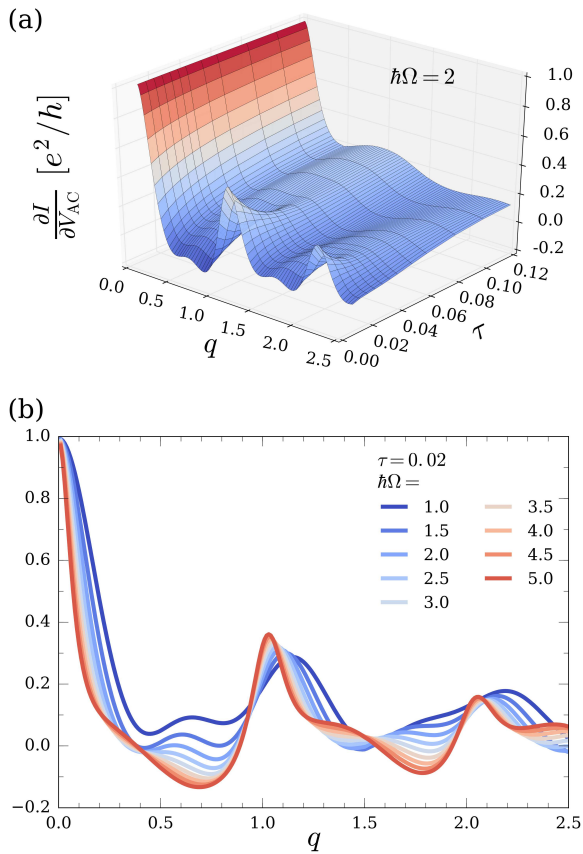}
\caption{
  (Color online)
  (a) Three-dimensional plot of the differential conductance $\partial I / \partial V_{\rm AC}$ with $\hbar\Omega = 2$ for various values of $q$ and $\tau$.
  (b) Cut of the differential conductance at $\tau=0.02$ for various values of $\Omega$.
  The dot parameters are taken as $\Gamma_{L,R}=1$, $\epsilon_{d}=-3$, and $U=6$ with the inverse temperature $\beta=100$.
  }
\label{fig: Conductance}
\end{figure}

The coexistence of the leviton and the Kondo resonance results in enhancement of the dc current
\begin{equation}
  I=\frac{e}{\hbar} 
  \int d\omega
       {\rm Tr}  \left[
       \left(\frac{-1}{\pi}{\rm Im} \bm{G}^{r}(\omega) \right) 
       \left( \bm{f}_{L}(\omega) - \bm{f}_{R}(\omega) \right) 
       \right],
\end{equation}
where the trace is taken over the Floquet indices.
The dependence of the differential conductance $\partial I / \partial V_{\rm AC}$ on $q$ is shown in FIG.~\ref{fig: Conductance}(a) for various values of $\tau$.
At $\tau=0.01$, the conductance is repeatedly enhanced around $q \in\mathbb{Z}$, indicating the reformation of the many-body resonance state through which the leviton flows.
The peak height around positive integer values of $q=n$ is crucially dependent on the width $\tau$ of the Lorentzian pulses because of the weight factor $|u_{-n}|^2=e^{-4\pi n\tau}$.
The unitarity limit is reached in the delta-pulse limit $\tau\rightarrow 0$, where the system is essentially in equilibrium due to the gauge structure of the many-electron system.
Hence, the characteristic $q$ dependence of the leviton transport results from the recovery of the equilibrium Kondo resonance under the sharp Lorentzian pulses.
The opposite large $\tau$ limit corresponds to a stationary voltage-biased QD because the Lorentzian-shaped periodic pulses are reduced to the dc voltage with amplitude $V_{\rm AC}$.
In the dc limit, the Kondo resonance vanishes for $q> k_{B}T_{K}/\hbar\Omega \simeq 0.082$ in contrast to the corresponding sharp-pulse cases.

The cut of the differential conductance $\partial I / \partial V_{\rm AC}$ at $\tau = 0.02$ is shown in FIG.~\ref{fig: Conductance}(b) for various values of $\Omega$.
The peaks around $q \in\mathbb{Z}$ become clear for large $\Omega$ cases, where
the transport channel through the Kondo peak is isolated from the other Floquet sidebands located at intervals of $\hbar\Omega$.
The differential conductance shows rich transport properties away from the optimal points as well.
The peak around $q \simeq 0.5$ for $\hbar\Omega=1$ results from the photon-assisted processes where electrons tunnel through the energy level at $\epsilon_{d}+Un_{d}$ by absorbing photons.
The peak is suppressed and become negative for large values of $\Omega$ due to the hole contributions generated by the non-quantized Lorentzian pulses.
The large dc offset of the Lorentzian pulses with $\hbar\Omega=5$ cause additional peaks around $q\simeq 1.4$ and $2.4$.
The various peak structures of different origin can be distinguished by their characteristic $\Omega$ dependence.
In particular, the enhancement of the conductance around $q \in \mathbb{Z}$ may provide a future experimental evidence of the leviton tunneling through the Kondo resonance.
Recent experiments on levitons \cite{dubois2013minimal,jullien2014quantum} have succeeded in producing Lorentzian-shaped pulse of $\Omega=6$GHz with $\tau=0.09$ at electron temperature $T_{e}\simeq 35$mK.
It seems necessary to generate sharper and higher-frequency Lorentzian pulses to observe the repeated emergence of the Kondo resonance of the typical energy scale $T_{K}\simeq 0.7$K$\sim 15$GHz in a Kondo QD system \cite{PhysRevLett.106.176601}.

In conclusion, we have demonstrated the coherent transport of levitons through the Kondo resonance realized in a QD system.
The dynamical formation of the Kondo resonance and its coexistence with the levitons can be identified as the enhancement of the dc transport under quantized Lorentzian pulses.
Since the leviton carries rich information on the many-body resonance state,
we can probe the dynamical properties of the interacting system by measuring the quantum interference and the noise spectroscopy of the leviton.
The present study also opens new possibilities for designing a quasiparticle excitation in interacting electron systems by engineering a time-dependent field.



T.J.S. acknowledges 
Taichi Hinokihara, 
Adrien Bolens, 
Rui Sakano, 
and Seiji Miyashita for fruitful discussions and comments.
T.J.S. is supported by Advanced Leading Graduate Course for Photon Science (ALPS).

\bibliography{revised_reference}

\end{document}